\documentstyle{article}[8pt]

\begin{document}
\begin{center}
{\Large {\bf Virial relations for the Dirac equation and their 
applications
to calculations of H-like atoms
}}
\end {center}
%
%
%
%
%
\begin{center}
{\large V. M. Shabaev}
\end{center}

%
%
%
\begin{center}
{{\it Department of Physics, St. Petersburg    
State University},\\ {\it Oulianovskaya Street 1, Petrodvorets,    
St. Petersburg 198504, Russia}}
\end{center}

\label{c_shab}

\begin{abstract}

Virial relations for the Dirac equation in a central field
and their applications to calculations of H-like atoms
are considered.  It is demonstrated that using these relations
allows one to evaluate various average values for a hydrogenlike
atom. The corresponding relations 
for  non-diagonal matrix elements provide 
an effective method
for analytical evaluations of infinite sums 
that occur in calculations based on using the reduced 
Coulomb-Green function. 
In particular, this method can
be used for calculations of higher-order corrections
to the hyperfine splitting and to the $g$ factor in 
hydrogenlike atoms.

\end{abstract}


\section{Introduction}    

\index{Virial relations|(}

In non-relativistic quantum mechanics, the virial theorem
for a particle moving in a central field $V(r)$ is given by
the well known equation:
\begin{eqnarray} \label{06nonrel}
\langle T \rangle = \frac{1}{2}\langle r\frac{dV}{dr}\rangle\,,
\end{eqnarray}   
where $\langle T \rangle$ denotes the average value of the kinetic
energy in a stationary state.
This theorem can easily be derived from the equation 
\cite{06hirschfelder60}
\begin{eqnarray} \label{06hir}
\frac{d}{dt}\langle A \rangle=\frac{i}{\hbar}\langle [H,A] \rangle =0\,,
\end{eqnarray}
if we take $A=({\bf r}\cdot{\bf p})$.
 Equation (\ref{06hir}), which is generally called as the
hypervirial theorem, 
can also be employed to derive virial relations for
diagonal matrix elements of other operators 
\cite{06epstein62}.
An extention of these relations to the case of non-diagonal
matrix elements was considered in \cite{06epstein67,06blanchard74}.

For the Dirac equation in a central field,
the virial theorem was first derived
by Fock \cite{06fock30}. If we denote the bound-state
energy by $E$, it gives
\begin{eqnarray} \label{06fock}
E=\langle mc^2 \beta\rangle +\langle r\frac{dV}{dr}\rangle
+\langle V \rangle\,,
\end{eqnarray}
where $\beta$ is the Dirac matrix.
For the Coulomb field, one easily finds
\begin{eqnarray}
E=\langle mc^2 \beta\rangle\,.
\end{eqnarray}
Some additional virial relations for the Dirac equation
were obtained by a number of authors (see, e.g., 
\cite{06epstein62,06goldman82}
and references therein).
Virial relations,
which yield recurrence formulas for various average values, 
were obtained in 
\cite{06epstein62,06shabaev84,06vrscay88,06shabaev91}.
In the case of the Coulomb field, these relations can be employed
to derive explicit formulas for the average values
$\langle r^s \rangle$, $\langle r^s\beta \rangle$, and
$\langle i r^s (\mbox{\boldmath $\alpha$}\cdot {\bf n})\beta\rangle$,
where $\mbox{\boldmath $\alpha$}$
 is a vector incorporating the Dirac matrices,
${\bf n}={\bf r}/r$, and $s$ is integer.
The corresponding recurrence relations
for non-diagonal matrix elements were  derived in \cite{06shabaev91}. 
In the case of the Coulomb field,
it was found that these relations can be employed to
derive explicit formulas for the first-order corrections
to the Dirac wave function due to interaction with  
perturbative potentials of the form $\sim r^s$, $r^s\beta$,  
$ r^s (\mbox{\boldmath $\alpha$}\cdot {\bf n})$, and
$ir^s (\mbox{\boldmath $\alpha$}\cdot {\bf n})\beta$.
Later on (see references below), 
this method was used for calculations
of various corrections to the energy levels, 
to the hyperfine structure splitting, 
and to the bound-electron $g$ factor.
In constract to direct analytical and numerical calculations,
the virial relation method allows one to derive formulas
for various physical quantities by simple algebraic
transformations.

In the present paper, following mainly to \cite{06shabaev91},
 we derive
the virial relations for the Dirac equation and examine
their applications
to calculations of H-like atoms.
Relativistic units ($\hbar=c=1$) are used in the paper.
 
\section{Derivation of the virial relations for the
Dirac equation}

For the case of a central field $V(r)$, the Dirac equation
has the form 
   \begin{eqnarray} \label{06direq}
 (-i\mbox{\boldmath $\alpha$}\cdot \mbox{\boldmath $\nabla$}
+\beta m+V({r}))\psi({\bf r})= 
 E \psi({\bf r})\,.
 \label{06dirac}
    \end{eqnarray}  
The wave function is conveniently represented by
\begin{eqnarray}
\psi({\bf r})=
\left(\begin{array}{c}
g(r)\Omega_{\kappa m}({\bf n})\\
if(r)\Omega_{-\kappa m}({\bf n})
\end{array}\right)\;,
\end{eqnarray}
where $\kappa=(-1)^{j+l+1/2}(j+1/2)$ is the quantum number determined
by the angular momentum and the parity of the state.
Substituting this expression into (\ref{06direq}), we obtain the radial
Dirac equations
\begin{eqnarray}
\frac{dG}{dr}+\frac{\kappa}{r}G-(E+m-V)F&=&0\,,\\
\frac{dF}{dr}-\frac{\kappa}{r}F+(E-m-V)G&=&0\,,
\end{eqnarray}
where $G(r)=rg(r)$ and $F(r)=rf(r)$.
Introducing the operator \cite{06drake81}
\begin{eqnarray}
H_{\kappa}=-i\sigma_y\frac{d}{dr}+\sigma_x\frac{\kappa}{r}
+\sigma_z m+V\,,
\end{eqnarray}
where $\sigma_x$, $\sigma_y$, and $\sigma_z$ are the Pauli
matrices, and denoting
\begin{eqnarray}
\phi({ r})=
\left(\begin{array}{c}
G(r)\\
F(r)
\end{array}\right)\;,
\end{eqnarray}
we obtain
\begin{eqnarray}
H_{\kappa}\phi=E\phi\,.
\end{eqnarray}
The operator $H_{\kappa}$ is self-adjoint in the space 
of two-component functions satisfying the boundary conditions
\begin{eqnarray}
\phi(0)=\phi(\infty)=0\,.
\end{eqnarray}
The scalar product in this space is defined by
\begin{eqnarray}
\langle a| b\rangle=\int_{0}^{\infty}dr\,(G_aG_b+F_aF_b)\,.
\end{eqnarray}
Let us denote the eigenvalues and the eigenvectors of the operator
$H_{\kappa}$ by $E_{n\kappa}$ and $\phi_{n\kappa}$, respectively,
where $n$ is the principal quantum number.
Taking into account the self-adjointness of  $H_{\kappa}$,
we can write down the following equations
\begin{eqnarray}\label{06com}
\langle n'\kappa'|(H_{\kappa'}Q-QH_{\kappa})|n\kappa\rangle&=&
(E_{n'\kappa'}-E_{n\kappa})
\langle n'\kappa'|Q|n\kappa\rangle\,,\\
\langle n'\kappa'|(H_{\kappa'}Q+QH_{\kappa})|n\kappa\rangle&=&
(E_{n'\kappa'}+E_{n\kappa})
\langle n'\kappa'|Q|n\kappa\rangle\,. \label{06anticom}
\end{eqnarray} 
Substituting $Q=r^s$, $i\sigma_y r^s$ into equation
(\ref{06com}) and
$Q=\sigma_zr^s$, $\sigma_x r^s$ into equation
(\ref{06anticom}), and using the commutation properties
of the Pauli matrices, we obtain \cite{06shabaev91}
\begin{eqnarray}\label{06bas1}
&(E_{n'\kappa'}-E_{n\kappa})\langle n'\kappa'|r^s|n\kappa\rangle
=-s\langle n'\kappa'|i\sigma_yr^{s-1}|n\kappa\rangle\nonumber\\
&\;\;\;\;\;\;\;\;\;\;\;\;\;\;\;\;\;\;\;
+(\kappa'-\kappa)\langle n'\kappa'|\sigma_x r^{s-1}|n\kappa\rangle\,,\\
&(E_{n'\kappa'}-E_{n\kappa})\langle n'\kappa'|i\sigma_yr^s|n\kappa\rangle
=s\langle n'\kappa'|r^{s-1}|n\kappa\rangle\nonumber\\
&\;\;\;\;\;\;\;\;\;\;\;\;\;\;\;\;\;\;\;
-(\kappa'+\kappa)\langle n'\kappa'|\sigma_z r^{s-1}|n\kappa\rangle
+2m\langle n'\kappa'|\sigma_x r^{s}|n\kappa\rangle\,, 
\label{06bas2}\\
&(E_{n'\kappa'}+E_{n\kappa})\langle n'\kappa'|\sigma_zr^s|n\kappa\rangle
=s\langle n'\kappa'|\sigma_xr^{s-1}|n\kappa\rangle\nonumber\\
&\;\;\;\;\;\;\;\;\;\;\;\;\;\;\;\;\;\;\;
-(\kappa'-\kappa)\langle n'\kappa'|i\sigma_y r^{s-1}|n\kappa\rangle
+2m\langle n'\kappa'|r^{s}|n\kappa\rangle \nonumber\\
&+2\langle n'\kappa'|\sigma_zVr^{s}|n\kappa\rangle\,,
\label{06bas3}\\
&(E_{n'\kappa'}+E_{n\kappa})\langle n'\kappa'|\sigma_xr^s|n\kappa\rangle
=-s\langle n'\kappa'|\sigma_zr^{s-1}|n\kappa\rangle\nonumber\\
&\;\;\;\;\;\;\;\;\;\;\;\;\;\;\;\;\;\;\;
+(\kappa'+\kappa)\langle n'\kappa'|r^{s-1}|n\kappa\rangle
+2\langle n'\kappa'|\sigma_xVr^{s}|n\kappa\rangle\,. \label{06bas4}
\end{eqnarray}
In the next sections, we apply these equations 
for calculations of the average values of various
physical quantities as well as for calculations of various
higher-order corrections.

\section{Application of the virial relations for
evaluation of the average values}

Let consider equations (\ref{06bas1})-(\ref{06bas4})
for the Coulomb field
($V(r)=-\alpha Z/r$)
and for diagonal matrix elements ($n'\kappa'=n\kappa$).
Denoting
\begin{eqnarray} \label{06defaa}
A^s&=&\int_{0}^{\infty}dr\; r^s(G_{n\kappa}^2+F_{n\kappa}^2)\,,\\
B^s&=&\int_{0}^{\infty}dr\; r^s(G_{n\kappa}^2-F_{n\kappa}^2)\,,
\label{06defbb}\\
C^s&=&2\int_{0}^{\infty}dr\; r^sG_{n\kappa}F_{n\kappa}\,,
\label{06defcc}
\end{eqnarray}
we obtain \cite{06epstein62,06shabaev84,06shabaev91}
\begin{eqnarray}\label{06rec1}
2mA^s-2E_{n\kappa}B^s&=&2\alpha Z B^{s-1}-sC^{s-1}\,,\\
2mC^s&=&-sA^{s-1}+2\kappa B^{s-1}\,, \label{06rec2} \\
2E_{n\kappa} C^s&=&2\kappa A^{s-1}-sB^{s-1}-2\alpha ZC^{s-1}\,.
\label{06rec3}
\end{eqnarray}
From these equations, one easily finds
\begin{eqnarray} \label{06rec4}
-[(s+1)E_{n\kappa}+2\kappa m]A^s+[(s+1)m+2\kappa E_{n\kappa}]B^s
+2\alpha Zm C^s=0
\end{eqnarray}
Using equations (\ref{06rec1})-(\ref{06rec4}) for 
$s=0,1$, we obtain  
\begin{eqnarray} \label{06b0}
B^0=\frac{E_{n\kappa}}{m}\,,\;\;\;\;\;
C^0=\frac{\kappa}{\alpha Z}
\frac{m^2-E_{n\kappa}^2}{m^2}\,,\;\;\;\;\;
B^{-1}=\frac{m}{\alpha Z}\frac{m^2-E_{n\kappa}^2}{m^2}\,.
\end{eqnarray}
In addition, according to the Hellmann-Feynman theorem, we have 
\begin{eqnarray}
\frac{\partial E_{n\kappa}}{\partial\kappa}=
\langle n \kappa |\frac{\partial H_{\kappa}}
{\partial\kappa}|n\kappa\rangle=
\langle n \kappa |\frac{\sigma_x}
{r}|n\kappa\rangle\,.
\end{eqnarray}
It yields
\begin{eqnarray} \label{06cm1}
C^{-1}=
\frac{\partial E_{n\kappa}}
{\partial \kappa}=\frac{(\alpha Z)^2\kappa m}{N^3\gamma}\,,
\end{eqnarray}
where $N=\sqrt{(\gamma+n_r)^2+(\alpha Z)^2}$, 
$\gamma=\sqrt{\kappa^2-(\alpha Z)^2}$, and $n_r=n-|\kappa|$.
The derivative with respect 
to $\kappa$ in equation (\ref{06cm1})
must be taken at a fixed $n_r$.
Using the formulas for $B^0$, $C^0$, $B^{-1}$, and $C^{-1}$
given above
and reccurence equations (\ref{06rec1})-(\ref{06rec4}),
we can calculate the integrals
 $A^s$, $B^s$, and $C^s$  for any integer $s$.
Explicit formulas for these calculations were derived
in \cite{06shabaev84}. The formulas expressing the
integrals  $A^{s+1}$, $B^{s+1}$, and $C^{s+1}$
in terms of the integrals  $A^s$, $B^s$, and $C^s$ are
\begin{eqnarray}
A^{s+1}&=&\{2\alpha Z E_{n\kappa}(s+1)A^s+2\alpha Z m(s+2)B^s\nonumber\\
&&-(s+1)[sm+2(m+\kappa E_{n\kappa})]C^s\}
\{2(s+2)(m^2-E_{n\kappa}^2)\}^{-1}\,,\\
B^{s+1}&=&\{2\alpha Z m(s+1)A^s+
2\alpha Z E_{n\kappa}(s+2)B^s\nonumber\\
&&-(s+1)[2\kappa m+(s+2)E_{n\kappa}]C^s\}
\{2(s+2)(m^2-E_{n\kappa}^2)\}^{-1}\,,\\
C^{s+1}&=&\frac{1}{2m}[2\kappa B^s-(s+1)A^s]\,.
\end{eqnarray}  
The reversed formulas are
\begin{eqnarray}
A^s&=&\frac{4\alpha Z(s+2)(mB^{s+1}-E_{n\kappa}A^{s+1})}
{(s+1)[(s+1)^2-4\gamma^2]}+ 
\frac{(s+1)m+2\kappa E_{n\kappa}}
{\alpha Z m[(s+1)^2-4\gamma^2]}\nonumber\\
&&\times \{[(s+2)m+2\kappa E_{n\kappa}]B^{s+1}
-[(s+2)E_{n\kappa}+2\kappa m]A^{s+1}\}\,,\\
B^s&=&\frac{4\alpha Z(mA^{s+1}-E_{n\kappa}B^{s+1})}
{(s+1)^2-4\gamma^2}- 
\frac{(s+1) E_{n\kappa}+2\kappa m}
{\alpha Z m[(s+1)^2-4\gamma^2]}\nonumber\\
&&\times \{[(s+2)E_{n\kappa}+2\kappa m]A^{s+1}
-[(s+2)m+2\kappa E_{n\kappa}]B^{s+1}\}\,,\\
C^s&=&\frac{1}{2\alpha Z m}[(s+1)E_{n\kappa}+2\kappa m]A^s
-\frac{1}{2\alpha Z m}[(s+1)m+2\kappa E_{n\kappa}]B^s\,.
\end{eqnarray}
Employing these formulas, one easily
finds
\begin{eqnarray}\label{06c1}
C^1&=&\frac{2\kappa E_{n\kappa}-m}{2m^2}\,,\\
A^{-1}&=&\frac{\alpha Z m}{\gamma N^3}(\kappa^2+n_r\gamma)\,,\\
B^{-1}&=&\frac{m^2-E_{n\kappa}^2}{\alpha Z m}\,,\\
A^{-2}&=&\frac{2(\alpha Z)^2\kappa 
[2\kappa(\gamma+n_r)-N]m^2}
{N^4(4\gamma^2-1)\gamma}\,,\\
B^{-2}&=&\frac{2(\alpha Z)^2[2\gamma^2N-\kappa(\gamma+n_r)]m^2}
{N^4(4\gamma^2-1)\gamma}\,,\\
C^{-2}&=&\frac{2(\alpha Z)^3[2\kappa(\gamma+n_r)-N]m^2}
{N^4(4\gamma^2-1)\gamma}\,,
\label{06cm2}\\
A^{-3}&=&\frac{2(\alpha Z)^3 m^3}
{N^5(4\gamma^2-1)\gamma(\gamma^2-1)}\nonumber\\
&&\times[N^2(1+2\gamma^2)-3\kappa N(\gamma +n_r)
+3(\alpha Z)^2(N^2-\kappa^2)]
\,.
\end{eqnarray}
It should be noted here that the integral $A^{-3}$ exists
only for $|\kappa|\ge 2$.
The integral $C^1$ occurs in calculations of the
bound-electron $g$ factor. The integrals 
$C^{-2}$ and $A^{-3}$ occur in calculations of 
the magnetic dipole and electric quadrupole hyperfine
splitting, respectively, (see \cite{06shabaev94} for details).
Formulas (\ref{06b0})-(\ref{06cm1}) and
(\ref{06c1})-(\ref{06cm2}) were also employed
in calculations of the recoil corrections to the atomic
energy levels \cite{06shabaev85,06shabaev94a}.

\section{Application of the virial relations for calculations
of higher-order corrections}

In calculations of higher-order corrections to various
physical quantities one needs to evaluate the sums
\begin{eqnarray}\label{06sumdef}
|i,s,\kappa',n\kappa\rangle\equiv
\sum_{n'}^{(E_{n'\kappa'}\ne E_{n\kappa})}
\frac{|n'\kappa'\rangle \langle n'\kappa'|R_i^s|n\kappa\rangle}
{E_{n\kappa} -E_{n'\kappa'}}\,,
\end{eqnarray}
where $R_1^s=r^s$, $R_2^s=\sigma_z r^s$,
$R_3^s=\sigma_x r^s$, and $R_4^s=i\sigma_y r^s$.
For instance, to derive the first-order correction
to the hydrogenic wave function due to the magnetic dipole 
 hyperfine interaction, we need to evaluate the expression
 (\ref{06sumdef}) for $R_3^{-2}$. Let us consider how
virial relations (\ref{06bas1})-(\ref{06bas4}) can be
employed for calculations of these sums in the case
of a hydrogenlike atom ($V(r)=-\alpha Z/r$).
For this case,  equations (\ref{06bas1})-(\ref{06bas4})
can be rewritten in the following form
\begin{eqnarray}\label{06basc1}
(E_{n\kappa}-E_{n'\kappa'})A_{n'\kappa',n\kappa}^s&=&
sD_{n'\kappa',n\kappa}^{s-1}+(\kappa-\kappa')C_{n'\kappa',n\kappa}^{s-1}\,,\\
(E_{n\kappa}-E_{n'\kappa'})D_{n'\kappa',n\kappa}^s&=&
-2mC_{n'\kappa',n\kappa}^{s}-sA_{n'\kappa',n\kappa}^{s-1}
+(\kappa'+\kappa)B_{n'\kappa',n\kappa}^{s-1}\,, 
\label{06basc2}\\
(E_{n\kappa}-E_{n'\kappa'})B_{n'\kappa',n\kappa}^s&=&
-2mA_{n'\kappa',n\kappa}^{s}+2E_{n\kappa}B_{n'\kappa',n\kappa}^{s}
-sC_{n'\kappa',n\kappa}^{s-1}\nonumber \\
&&+(\kappa' -\kappa)D_{n'\kappa',n\kappa}^{s-1}
+2\alpha ZB_{n'\kappa',n\kappa}^{s-1}\,, 
\label{06basc3}\\
(E_{n\kappa}-E_{n'\kappa'})C_{n'\kappa',n\kappa}^s&=&
2E_{n\kappa}C_{n'\kappa',n\kappa}^{s}
-(\kappa'+\kappa)A_{n'\kappa',n\kappa}^{s-1}
+sB_{n'\kappa',n\kappa}^{s-1}\nonumber \\
&&+2\alpha ZC_{n'\kappa',n\kappa}^{s-1}\,, \label{06basc4}
\end{eqnarray}
where
\begin{eqnarray}
A_{n'\kappa',n\kappa}^s&=&\int_{0}^{\infty}dr\; r^s
(G_{n'\kappa'}G_{n\kappa}+F_{n'\kappa'}F_{n\kappa})\,,\\
B_{n'\kappa',n\kappa}^s&=&\int_{0}^{\infty}dr\; r^s
(G_{n'\kappa'}G_{n\kappa}-F_{n'\kappa'}F_{n\kappa})\,,\\
C_{n'\kappa',n\kappa}^s&=&\int_{0}^{\infty}dr\; r^s
(G_{n'\kappa'}F_{n\kappa}+F_{n'\kappa'}G_{n\kappa})\,,\\
D_{n'\kappa',n\kappa}^s&=&\int_{0}^{\infty}dr\; r^s
(G_{n'\kappa'}F_{n\kappa}-F_{n'\kappa'}G_{n\kappa})\,.
\end{eqnarray}
From equations (\ref{06basc1})-(\ref{06basc4}), we obtain
\begin{eqnarray} \label{06basc5}
\lefteqn{(E_{n\kappa}-E_{n'\kappa'})(E_{n\kappa}D_{n'\kappa',n\kappa}^{s}
+mC_{n'\kappa',n\kappa}^{s})}\nonumber\\
&&=[-sE_{n\kappa}-m(\kappa'+\kappa)]
A_{n'\kappa',n\kappa}^{s-1}
+[sm +E_{n\kappa}(\kappa'+\kappa)]
B_{n'\kappa',n\kappa}^{s-1}\nonumber\\
&&\;\;\;+2\alpha Z m C_{n'\kappa',n\kappa}^{s-1}\,.
\end{eqnarray}
For $n\kappa=n'\kappa'$, this equation turns into equation
(\ref{06rec4}). 

Let us consider first the case $\kappa=\kappa'$.
Taking into account that  
$A_{n'\kappa,n\kappa}^0=\delta_{n'n}$, we obtain
\begin{eqnarray}
B_{n'\kappa,n\kappa}^0&=&\frac{1}{m}
(E_{n\kappa}-E_{n'\kappa})(E_{n\kappa}D_{n'\kappa,n\kappa}^{1}
+mC_{n'\kappa,n\kappa}^{1}\nonumber\\
&&+\alpha Z D_{n'\kappa,n\kappa}^{0}
-\kappa B_{n'\kappa,n\kappa}^{0})
+\frac{E_{n\kappa}}{m}\delta_{nn'}\,.
\end{eqnarray}
Multiplying this equation with $|n'\kappa\rangle$
and summing over $n'$, we derive
\begin{eqnarray} \label{06sumb0}
|2,0,\kappa,n\kappa\rangle&=&\frac{1}{m}
(I-|n\kappa\rangle \langle n\kappa|)\nonumber\\
&&\times(E_{n\kappa}i\sigma_y r
+m\sigma_x r+\alpha Z i\sigma_y -\kappa\sigma_z )
|n\kappa\rangle\,,
\end{eqnarray}
where $I$ is the identity operator.
From equations (\ref{06basc5}) and  (\ref{06sumb0}),
we can derive the sum $|3,0,\kappa,n\kappa\rangle$.
Then, using equations (\ref{06basc2})-(\ref{06basc4}) and
(\ref{06basc5}),
we can calculate all the sums $|i,s,\kappa,n\kappa\rangle$
for $i=1,2,3$ and $s=0,1,2,...$ . In particular, for the sum
$|3,1,\kappa,n\kappa\rangle$ that occurs in calculations of
various corrections to the bound-electron $g$ factor, we
 find
\begin{eqnarray}\label{06gfac}
|3,1,\kappa,n\kappa\rangle &=&\frac{\kappa}{m^2}
(I-|n\kappa\rangle \langle n\kappa|)\nonumber\\
&&\times \Bigl[\Bigl
(E_{n\kappa}-\frac{m}{2\kappa}\Bigr)ri\sigma_y+mr\sigma_x+
\alpha Z i\sigma_y-\kappa \sigma_z\Bigr]|n\kappa\rangle\,.
\end{eqnarray}
The sums $|4,s,\kappa,n\kappa\rangle$ 
for $s\ne -1$ can be calculated by employing equation (\ref{06basc1}).
The sum  $|4,-1,\kappa,n\kappa\rangle$ is easily derived from
the relation
\begin{eqnarray}
D_{n'\kappa,n\kappa}^{-1}=
(E_{n\kappa}-E_{n'\kappa})\langle n'\kappa|{\rm ln}r|n\kappa\rangle\,,
\end{eqnarray}
which is obtained by
differentiation of equation (\ref{06basc1}) with respect to $s$.

Let us consider now the case $\kappa'\ne \kappa$.
From equation (\ref{06basc1}) we find
\begin{eqnarray} \label{06sumdm1}
|3,-1,\kappa',n\kappa\rangle=\frac{1}{\kappa-\kappa'}
|n\kappa\rangle \,.
\end{eqnarray}
Then, using equations (\ref{06basc1})-(\ref{06basc4}) and
(\ref{06basc5}),
we can calculate all the sums $|i,s,\kappa',n\kappa\rangle$
for $i=1,2,3,4$ and $s=-2,-3,-4,...$ (if, of course, the 
corresponding sum exists). 
For the sum $|3,-2,\kappa',n\kappa\rangle$ that occurs
in calculations of various corrections to the hyperfine
splitting, we find
\begin{eqnarray} \label{06hfs1}
|3,-2,\kappa',n\kappa\rangle
&=&\{[1-(\kappa-\kappa ')^{2}][1-(\kappa+\kappa ')^{2}]+4(\alpha Z)^{2}\}
^{-1}\nonumber\\
& &\times\Bigl[[1-(\kappa+\kappa  ')^{2}]\Bigl(\frac{4\alpha Z m}{\kappa^{2}
-\kappa '^{2}}+(\kappa '-\kappa)r^{-1}+r^{-1}\sigma_{z}\nonumber\\
& &-\frac{2}{\kappa+\kappa '}(m\sigma_{x}+E_{n\kappa}
i\sigma_{y})\Bigl)+\frac{4\alpha Z[(\kappa '+\kappa)m-E_{n\kappa}]}
{\kappa-\kappa '}\nonumber\\
& & +2\alpha Z[\sigma_{x}r^{-1}+(\kappa '+\kappa)r^{-1}i\sigma_{y}]
\Bigr]|n\kappa\rangle\;.
\end{eqnarray}
For $\kappa=\pm\kappa'$, the corresponding sums
can be calculated by taking the limit $\kappa' \rightarrow
\pm \kappa$.  So, taking the limit $\kappa' \rightarrow
 \kappa$ in equation  (\ref{06basc1}), we obtain
 \begin{eqnarray} \label{06sumdm1a}
|3,-1,\kappa,n\kappa\rangle=\frac{\partial}{\partial\kappa}
|n\kappa\rangle \,,
\end{eqnarray}
where, as in (\ref{06cm1}), the derivative with respect 
to $\kappa$ must be taken at a fixed $n_r$.
Then, the other sums with $\kappa'=\kappa$ and $s=-1,-2,-3,...$
can be calculated by using equations
(\ref{06basc1})-(\ref{06basc4}) and (\ref{06basc5}). 
In particular, we obtain
\begin{eqnarray} \label{06hfs2}
|3,-2,\kappa,n\kappa\rangle
&=&\frac{1}{4(\alpha Z)^{2}+(1-4\kappa^{2})}\Bigl[2\alpha Z\sigma_{x}
r^{-1}+4\alpha Z\kappa i \sigma_{y}r^{-1}\nonumber\\
&&+(1-4\kappa^{2})\sigma_{z}r^{-1}-\frac{(1-4\kappa^{2})}{\kappa}(
E_{n\kappa}i\sigma_{y}+m\sigma_{x})\nonumber\\
&&-\frac{2(\alpha Z)^{3}\kappa m}{N^{3}\gamma}\Bigr]|n\kappa\rangle-
\frac{2\alpha Z(2E_{n\kappa}-\frac{m}{\kappa})}
{4(\alpha Z)^{2}+(1-4\kappa^{2})}\frac{\partial}
{\partial\kappa}|n\kappa\rangle\;.
\end{eqnarray}
The case  $\kappa'=-\kappa$
can be considered in the same way.

Concluding this section, we give the explicit formulas
for $\frac{\partial}{\partial\kappa}
|n\kappa\rangle$ for the $1s$ and $2s$ states
 \cite{06shabaeva95}.
 For the $1s$ state:
\begin{eqnarray}
\frac{\partial}{\partial\kappa}|n\kappa\rangle=
\left( \begin{array}{c}
\tilde{G}_{1s}(r)\\
\tilde{F}_{1s}(r)
\end{array} \right)\;,
\end{eqnarray}
where
\begin{eqnarray}
\tilde{G}_{1s}(r)&=&\frac{k}{\sqrt{1-\gamma}}\exp{(-t/2)}t^{\gamma}
\Bigl(\frac{\psi(2\gamma+1)}{\gamma}+(\gamma+1)\nonumber\\
&&-\frac{1}{2\gamma}
-\frac{t}{2}-\frac{1}{\gamma}\ln{t}\Bigr)\;,\\
\tilde{F}_{1s}(r)&=&-\frac{k}{\sqrt{1+\gamma}}\exp{(-t/2)}t^{\gamma}
\Bigl(\frac{\psi(2\gamma+1)}{\gamma}+(\gamma+1)\nonumber\\
&&+\frac{1}{2\gamma}
-\frac{t}{2}-\frac{1}{\gamma}\ln{t}\Bigr)\;,
\end{eqnarray}
$$
t=\frac{2\alpha Z m r}{N}\;,\;\;\; k=\frac{(2\alpha Z)^{\frac{3}{2}}
m^{\frac{1}{2}}}
{2\sqrt{2\Gamma(2\gamma+1)}}\;,
$$
$\Gamma(x)$ is the gamma-function, and $\psi(x)=\frac{d}{dx}\ln\Gamma(x)$.
For the $2s$ state:
\begin{eqnarray}
\frac{\partial}{\partial\kappa}|n\kappa\rangle=
\left( \begin{array}{c}
\tilde{G}_{2s}(r)\\
\tilde{F}_{2s}(r)
\end{array} \right)\;,
\end{eqnarray}
where
\begin{eqnarray}
\tilde{G}_{2s}(r)
&=&k'\exp{(-\frac{t}{2})}t^{\gamma}\frac{\sqrt{2+N}}{N^{2}-2}
\Bigl\{\frac{t^{2}}{2(N-1)}\nonumber\\
&&-\Bigl(\frac{2N^{4}-4N^{3}+5N^{2}-3N+2}
{2N(N-1)^{2}}
+\frac{2\psi(2\gamma+1)}{N-1}\Bigr)t\nonumber\\
&&+\frac{N^{4}-2N^{3}+N-2}
{2(N-1)}+2N\psi(2\gamma+1)\nonumber\\
&&+\frac{2}{N-1}t\ln{t}-2N\ln{t}\Bigr\}\;,\\
\tilde{F}_{2s}(r)
&=&-k'\exp{(-\frac{t}{2})}t^{\gamma}\frac{\sqrt{2-N}}{N^{2}-2}
\Bigl\{\frac{t^{2}}{2(N-1)}\nonumber\\
&&-\Bigl(\frac{2N^{4}-2N^{3}+N^{2}+3N-2}
{2N(N-1)^{2}}
+\frac{2\psi(2\gamma+1)}{N-1}\Bigr)t\nonumber\\
&&+\frac{N^{5}+5N^{2}-8N-4}
{2N(N-1)}+2(N+2)\psi(2\gamma+1)\nonumber\\
&&+\frac{2}{N-1}t\ln{t}-2(N+2)\ln{t}\Bigr\}\;,
\end{eqnarray}
$$
t=\frac{2\alpha Z m r}{N}\;,\;\;\;\;
k'=\frac{\sqrt{\Gamma(2\gamma+2)}}{\Gamma(2\gamma+1)}\frac{1}
{\sqrt{8N(N+1)}}\Bigl(\frac{2\alpha Z m}{N}\Bigr)^{\frac{1}{2}}\;.
$$

\section{Calculations of the bound-electron
$g$ factor and the hyperfine splitting in H-like atoms}

For the last few years a significant progress was achieved
in calculations of the bound-electron $g$ factor and the hyperfine
splitting in H-like atoms.
Formulas (\ref{06gfac}) and (\ref{06hfs2}) were extensively 
employed in these calculations.

In \cite{06shabaev01}, a complete $\alpha Z$-dependence
 formula for the recoil
correction of order $m/M$ to the $g$ factor of an H-like atom 
was derived. According to this formula,
which was confirmed by an independent derivation in 
\cite{06yelkhovsky01}, the recoil correction 
is given by ($e <0$)
\begin{eqnarray}
\Delta g&=&\frac{1}{\mu_0 m_a}\frac{i}{2\pi M}
\int_{-\infty}^{\infty} d\omega\;
\Biggl[\frac{\partial}{\partial {\cal H}}
\langle \tilde{a}|[p^k-D^k(\omega)+eA_{\rm cl}^k]
\nonumber\\
&&\times\tilde{G}(\omega+\tilde{E}_a)
[p^k-D^k(\omega)+eA_{\rm cl}^k]
|\tilde{a}\rangle
\Biggr]_{{\cal H}=0}\,.
\label{06recoilt}
\end{eqnarray}
Here $\mu_0$ is the Bohr magneton, $m_a$ is the angular momentum
projection of the state $a$,
 $p^k=-i\nabla^k$ is the momentum operator, 
${\bf A}_{\rm cl}=[\mbox{\boldmath ${\cal H}$}
\times {\bf r}]/2$ is the 
vector potential of the homogeneous
magnetic field $\mbox{\boldmath ${\cal H}$}$ 
directed along the $z$ axis,
$D^k(\omega)=-4\pi\alpha Z\alpha^l D^{lk}(\omega)$,
\begin{eqnarray} \label{06photon}
D^{il}(\omega,{\bf r})=-\frac{1}{4\pi}\Bigl\{\frac
{\exp{(i|\omega|r)}}{r}\delta_{il}+\nabla^{i}\nabla^{l}
\frac{(\exp{(i|\omega|r)}
-1)}{\omega^{2}r}\Bigr\}\,
\end{eqnarray}
is the transverse part of the photon propagator in the Coulomb 
gauge. The tilde sign indicates that the related quantity
(the wave function, the energy, and the Coulomb-Green function
$\tilde{G}(\omega)=\sum_{\tilde{n}}|\tilde{n}\rangle \langle 
\tilde{n}|[\omega-\tilde{E}_n(1-i0)]^{-1}$)
must be calculated  at the presence of the  homogeneous
magnetic field $\mbox{\boldmath ${\cal H}$}$
 directed along the $z$ axis.
In equation (\ref{06recoilt}) and below, 
the summation over the repeated indices 
($k=1,2,3$), which enumerate components of the three-dimensional
vectors, is implicit.  
For the practical calculations, this expression is conveniently
represented by the sum of the lower-order term and the higher-order
term, $\Delta g=\Delta g_{\rm L}+\Delta g_{\rm H}$, where 
\begin{eqnarray} \label{06recoil1}
\Delta g_{\rm L}&=&\frac{1}{\mu_0 m_a}\frac{1}{2M} 
\Biggl[\frac{\partial}{\partial {\cal H}}
\langle \tilde{a}|\Bigl\{
{\bf p}^2-\frac{\alpha Z}{r}[(\mbox{\boldmath $\alpha$}
\cdot {\bf p})
+(\mbox{\boldmath $\alpha$}\cdot {\bf n})
({\bf n}\cdot{\bf p})]\Bigr\}|\tilde{a}\rangle
 \Biggr]_{{\cal H}=0}\nonumber \\
&&-\frac{1}{m_a}
\frac{m}{M}\langle a|\Bigl([{\bf r}\times {\bf p}]_z
-\frac{\alpha Z}{2r}[{\bf r}\times \mbox{\boldmath $\alpha$}]_z
\Bigr)|a\rangle\,,\\
\Delta g_{\rm H}&=&\frac{1}{\mu_0 m_a}\frac{i}{2\pi M}
\int_{-\infty}^{\infty} d\omega\;
\Biggl[\frac{\partial}{\partial {\cal H}}
\langle \tilde{a}|\Bigl(D^k(\omega)-\frac{[p^k,V]}{\omega+i0}\Bigr)
\nonumber\\
&&\times\tilde{G}(\omega+\tilde{E}_a)
\Bigl(D^k(\omega)+\frac{[p^k,V]}{\omega+i0}\Bigr)|\tilde{a}\rangle
\Biggr]_{{\cal H}=0}\,,
\label{06recoil2}
\end{eqnarray}
where  $V(r)=-\alpha Z/r$ is the Coulomb
potential induced by the nucleus and
${\bf n}={\bf r}/r$.
The lower-order term can be calculated analytically by 
employing formula (\ref{06gfac}) and the formulas
for the average values presented above.
Let us consider this calculation in details \cite{06shabaev01}.
According to equation (\ref{06recoil1}), the lower-order term 
is the sum of two contributions,
 $\Delta g_{\rm L} = \Delta g_{\rm L}^{(1)} +\Delta g_{\rm L}^{(2)}$.
The first contribution is 
\begin{eqnarray} \label{06shabaeveq96}
\Delta g_{\rm L}^{(1)}=\frac{1}{\mu_0 m_a {\cal H}}
\frac{1}{M} \langle \delta a|
\Bigr[
{\bf p}^2-\frac{\alpha Z}{r}(\mbox{\boldmath $\alpha$}
\cdot{\bf p}
+(\mbox{\boldmath $\alpha$}
\cdot {\bf n})({\bf n}\cdot{\bf p})\Bigr]|a\rangle\,,
\end{eqnarray}
where $\delta a$ is the first-order correction to the electron
wave function due to interaction with the homogeneous magnetic field.
 Taking into account that 
${\bf p}^2=(\mbox{\boldmath $\alpha$} \cdot {\bf p})^2$ and
$(\mbox{\boldmath $\alpha$}\cdot {\bf p})
=H-\beta m-V$,
one easily obtains
\begin{eqnarray} \label{06shabaeveq97}
\langle \delta a|{\bf p}^2|a\rangle &=&\langle \delta a|
(E_a+\beta m-V)(E_a-\beta m-V)|a\rangle \nonumber\\
&&+i \langle \delta a|(\mbox{\boldmath $\alpha$}
\cdot \mbox{\boldmath $\nabla$} V)|a\rangle\,.
\end{eqnarray}
The second term in equation (\ref{06shabaeveq96}) can be transformed
as (see, e.g., \cite{06shabaev94a})
\begin{eqnarray} \label{06shabaeveq98}
-\langle \delta a|\frac{\alpha Z}{r}[\mbox{\boldmath $\alpha$}
\cdot {\bf p}
+(\mbox{\boldmath $\alpha$}
 \cdot {\bf n})({\bf n} \cdot {\bf p})]|a\rangle 
&=&-\langle \delta a|
\frac{\alpha Z}{r}\Bigr[2E_a-2\beta m
-2V\nonumber\\
&&+\frac{i}{r}(\mbox{\boldmath $\alpha$}
\cdot {\bf n})(\beta \kappa+1)
\Bigr]|a\rangle\,.
\end{eqnarray}
The wave function correction $|\delta a\rangle$ is defined by
\begin{eqnarray}
|\delta a\rangle&=&
\sum_n^{E_n\ne E_a}\frac{|n\rangle\langle n|
|e|\mbox{\boldmath $\alpha$}\cdot{\bf A}_{\rm cl} 
|a\rangle}
{E_a-E_n}\,.
\label{06shabaeveq71}
\end{eqnarray}
Since the operator sandwiched between $|a\rangle$ and 
$|\delta a\rangle$
in equation for $\Delta g_{\rm L}^{(1)}$ conserves
the angular quantum numbers, we need only that component of
 $|\delta a\rangle$ which has the same angular quantum numbers
as the unperturbed state  $|a\rangle$.
Using formula (\ref{06gfac}), we easily find
\begin{eqnarray} \label{06shabaeveq99}
|\delta a\rangle_{\kappa m_a}=
\left(\begin{array}{c}
X(r)\Omega_{\kappa m_a}({\bf n})\\
iY(r)\Omega_{-\kappa m_a}({\bf n})
\end{array}\right)\;,
\end{eqnarray}
where
\begin{eqnarray} \label{06shabaeveq100}
X(r)&=&b_0\Bigl\{\Bigl[\frac{2m\kappa -m+2\kappa E_a}{2m^2}r
+\frac{\alpha Z}{m^2}\kappa\Bigr]f(r)
+\frac{\kappa-2\kappa^2}{2m^2}g(r)\Bigr\}\,,\\
Y(r)&=&b_0\Bigl\{\Bigl[\frac{2m\kappa +m-2\kappa E_a}{2m^2}r
-\frac{\alpha Z}{m^2}\kappa\Bigr]g(r)
+\frac{\kappa+2\kappa^2}{2m^2}f(r)\Bigr\}\,,
\label{06shabaeveq101}\\
b_0&=&-\frac{e}{2}{\cal H}\frac{\kappa}{j(j+1)}m_a\,,
\label{06shabaeveq102}
\end{eqnarray}
$g(r)$ and $f(r)$ are the radial parts of the unperturbed
wave function defined above.
Integrating over the angular variables in 
equations (\ref{06shabaeveq97}) and (\ref{06shabaeveq98}), we find
\begin{eqnarray} \label{06shabaeveq104}
\Delta g_{\rm L}^{(1)}&=&\frac{\kappa}{j(j+1)}
\frac{m}{M}\int_{0}^{\infty}dr\;
r^2\Bigl\{X(r)g(r)[-2Vm-V^2+E_a^2-m^2]
\nonumber\\
&&+Y(r)f(r)[2Vm-V^2+E_a^2-m^2]\nonumber\\
&&+[X(r)f(r)+Y(r)g(r)]\frac{\alpha Z}{r^2}\kappa\Bigr\}\,.
\end{eqnarray}
Substituting expressions (\ref{06shabaeveq100}) and
 (\ref{06shabaeveq101})
into equation (\ref{06shabaeveq104}), we obtain
\begin{eqnarray} \label{06shabaeveq105}
\Delta g_{\rm L}^{(1)}&=&\frac{\kappa}{j(j+1)}
\frac{m}{M}\Bigl\{
\alpha Z\frac{2\kappa E_a-m}{m}C^0+
(\alpha Z)^2 \frac{\kappa}{m}C^{-1}\nonumber\\
&&+(E_a^2-m^2)\frac{\kappa}{m}C^{1}
+\alpha Z \frac{\kappa^2}{2m^2}C^{-2}+(E_a^2-m^2)
\frac{\kappa}{2m^2}A^0\nonumber\\
&&-\alpha Z\frac{\kappa^2}{m}A^{-1}
-(\alpha Z)^2\frac{\kappa}{2m^2}A^{-2}
-(E_a^2-m^2)\frac{\kappa^2}{m^2}B^0\nonumber\\
&&+\alpha Z\frac{3m\kappa-2\kappa^2 E_a}{2m^2}B^{-1}\Bigr\}\,.
\end{eqnarray}
Using the explicit expressions for
the integrals $A^s$, $B^s$, and $C^s$ given above,
 we obtain
\begin{eqnarray} \label{06shabaeveq108}
\Delta g_{\rm L}^{(1)}=\frac{m}{M}\,
\frac{\kappa^2}{2j(j+1)}\frac{m^2-E_a^2}{m^2}\,.
\end{eqnarray}
Consider now the contribution $\Delta g_{\rm L}^{(2)}$:
\begin{eqnarray} \label{06shabaeveq109}
\Delta g_{\rm L}^{(2)}=
-\frac{1}{m_a}\,\frac{m}{M}\,
\langle a|
\Bigl(l_z-\frac{\alpha Z}{2r}[{\bf r}\times 
\mbox{\boldmath $\alpha$}]_z\Bigr)|a\rangle\,.
\end{eqnarray}
Integrating over the angular variables and employing
the explicit expressions for the radial integrals derived above,
 we obtain
\begin{eqnarray} \label{06shabaeveq110}
\Delta g_{\rm L}^{(2)}&=&-\frac{m}{M}\,
\frac{1}{2j(j+1)}
\Bigl\{j(j+1)-\frac{3}{4}+l(l+1)\frac{m+E_a}{2m}
\nonumber\\
&&+(2j-l)(2j-l+1)\frac{m-E_a}{2m}
-\kappa^2\frac{m^2-E_a^2}{m^2}\Bigr\}\,.
\end{eqnarray}
For the sum $\Delta g_{\rm L}=\Delta g_{\rm L}^{(1)}+
\Delta g_{\rm L}^{(2)}$, we find
\begin{eqnarray} \label{06shabaeveq112}
\Delta g_{\rm L}=-\frac{m}{M}
\frac{2\kappa^2E_a^2+\kappa m E_a-m^2}{2m^2j(j+1)}\,.
\end{eqnarray}
To the two lowest orders in $\alpha Z$, we have
\begin{eqnarray} \label{06shabaeveq113}
\Delta g_{\rm L}=-\frac{m}{M}\,
\frac{1}{j(j+1)}\Bigl[\kappa^2+\frac{\kappa}{2}
-\frac{1}{2}-\Bigl(\kappa^2+\frac{\kappa}{4}\Bigr)
\frac{(\alpha Z)^2}{n^2}\Bigr]\,.
\end{eqnarray}
For the $1s$ state, formula (\ref{06shabaeveq112}) yields
\begin{eqnarray} \label{06shabaeveq114}
\Delta g_{\rm L}=\frac{m}{M}(\alpha Z)^2-\frac{m}{M}
\frac{(\alpha Z)^4}{3[1+\sqrt{1-(\alpha Z)^2}]^2}\,.
\end{eqnarray}
The first term in the right-hand side of this equation
reproduces the result of
\cite{06faustov70,06grotch70}.
The higher-order term $\Delta g_{\rm H}$
was evaluated numerically for the $1s$ state
in \cite{06shabaev02}. Formula  (\ref{06gfac}) was also
extensively used in that evaluation.

Let us consider now  the derivation of the 
nuclear-size correction to the $g$ factor of a low-$Z$ H-like atom
\cite{06glazov02}.
To find this correction, we have
to evaluate the expression
\begin{eqnarray}
  \Delta g=\frac{2}{\mu_0 m_a{\cal H}}
\sum_n^{n \ne a}\frac{\langle a |\delta V |
  n\rangle \langle n ||e|\mbox{\boldmath $\alpha$}\cdot{\bf A}_{\rm cl} 
  |a \rangle}{E_a-E_n}\,,
\end{eqnarray}
where $\delta V$ determines the deviation of the potential from
the pure Coulomb one.
Integrating over the angular variables, we obtain
\begin{eqnarray}
  \Delta g=
  \frac{2\kappa m}{j(j+1)}
  \sum_{n'}^{n'\ne n}
  \frac{\langle n \kappa | \delta V | n' \kappa \rangle
  \langle n' \kappa | r \sigma_x | n \kappa \rangle}
  {E_{n\kappa}-E_{n'\kappa}}\,,
\end{eqnarray}
where $|n\kappa\rangle$ are the two-component radial wave functions 
defined above.
Substituting expression (\ref{06gfac}) into this equation,
we get
\begin{eqnarray}
  \Delta g &=& \frac{2\kappa^2}{j(j+1)m}
\Bigl\{  \langle n \kappa | \delta V 
\left[\left(E_{n \kappa}-\frac{m}{2\kappa}\right)ri\sigma_y
  +mr\sigma_x+\alpha Zi\sigma_y-\kappa\sigma_z\right]
 | n \kappa \rangle \nonumber\\
&&-   \langle n \kappa | 
\left[\left(E_{n \kappa}-\frac{m}{2\kappa}\right)ri\sigma_y
  +mr\sigma_x+\alpha Zi\sigma_y-\kappa\sigma_z\right]
| n \kappa \rangle \nonumber\\
&&\times \langle n \kappa |\delta V | n \kappa \rangle
\Bigr\}\,.
\end{eqnarray}
We assume that the nuclear charge distribution is described by a spherically
symmetric density $\rho({\bf r})=\rho(r)$, which is normalized by the equation
\begin{eqnarray}
  \int d {\bf r}\; \rho({\bf r}) = 1\,.
\end{eqnarray}
The Poisson equation gives
\begin{eqnarray}
  \Delta(\delta V) ({\bf r})
 = 4\pi\alpha Z [\rho({\bf r}) - \delta ({\bf r})],
\end{eqnarray}
where $\Delta$ is the Laplacian.
When integrated with $\delta V$, the
 radial functions $g(r)$ and $f(r)$ can be
approximated by the lowest order term of the expansion in powers of $r$.
It follows that we have to evaluate the integral
\begin{eqnarray}
  I=\int_0^{\infty}\limits dr\; r^2 r^{2\gamma-2} \delta V\,.
\end{eqnarray}
Using the identity
\begin{eqnarray}
  r^{\beta}=\frac{1}{(\beta + 2)(\beta + 3)}\Delta r^{\beta+2}
\end{eqnarray}
and integrating by parts, we find
\begin{eqnarray}
  I =
  \int_0^{\infty}\limits dr ~ r^2 ~ \frac{1}{2\gamma(2\gamma+1)}
   ~ \Delta r^{2\gamma} ~ \delta V=
  \int_0^{\infty}\limits dr ~ r^2 ~ \frac{1}{2\gamma(2\gamma+1)}
   ~ r^{2\gamma} ~ \Delta (\delta V)
\nonumber\\
  = \frac{4\pi\alpha Z}{2\gamma(2\gamma+1)} \int_0^{\infty}\limits dr ~ r^2
  ~ r^{2\gamma} ~ \rho(r)=
  \frac{\alpha Z}{2\gamma(2\gamma+1)} ~ \langle r^{2\gamma} 
\rangle_{\rm nuc}\,,
\end{eqnarray}
where
\begin{eqnarray}
  \langle r^{2\gamma} \rangle_{\rm nuc} 
= \int d {\bf r}\; r^{2\gamma} \rho(r)\,.
\end{eqnarray}
For the correction to the $g$ factor we obtain
\begin{eqnarray}
  \Delta g &=& \frac{\kappa^2}{j(j+1)}
  \frac{\Gamma(2\gamma+1+n_r) 2^{2\gamma-1}}
  {\gamma(2\gamma+1)\Gamma^2(2\gamma+1) n_r!(N-\kappa)N^{2\gamma+2} }
\nonumber\\
&& \times \left\lbrack \left[ n_r^2 + (N-\kappa)^2 \right]
  \left( 1 - 2\kappa\frac{E_{n \kappa}}{m} \right) -
  2n_r(N-\kappa) \left( \frac{E_{n \kappa}}{m} - 2\kappa \right) \right\rbrack
\nonumber\\  
&&\times (\alpha Z)^{2\gamma+2} m^{2\gamma} \langle r^{2\gamma} 
\rangle_{\rm nuc} \,.
\label{06shabaevgl1}
\end{eqnarray}
For  $ns$-states, which are of particular interest,
the expansion of this expression to two lowest orders in $\alpha Z$ yields
\begin{eqnarray}
  \Delta g &=&\frac{8}{3n^3} (\alpha Z)^4 m^2 
\langle r^2 \rangle_{\rm nuc}
  \Biggl[ 1+(\alpha Z)^2
  \Biggl( \frac{1}{4} + \frac{12n^2-n-9}{4n^2(n+1)} 
\nonumber\\
&& + 2\psi(3) -\psi(2+n)
   - \frac{\langle r^2 \ln(2\alpha Zmr/n) \rangle_{\rm nuc}}
  {\langle r^2 \rangle_{\rm nuc}} \Biggr) \Biggr]\,,
\label{06shabaevgl2}
\end{eqnarray}
where $\psi(x)=\frac{d}{dx}\ln\Gamma(x)$.
For the $1s $ state, we have
\begin{eqnarray}
  \Delta g = \frac{8}{3} (\alpha Z)^4 m^2 \langle r^2 \rangle_{\rm nuc}
  \Biggl[ 1+(\alpha Z)^2
  \Biggl(2-C - \frac{\langle r^2 \ln(2\alpha Zmr) \rangle_{\rm nuc}}
  {\langle r^2 \rangle_{\rm nuc}} \Biggr) \Biggr]\,,
\label{06shabaevgl3}
\end{eqnarray}
where $C=0.57721566490\ldots$ is the Euler constant.
In the non-relativistic limit, we find
\begin{eqnarray}
 \Delta g = \frac{8}{3n^3} (\alpha Z)^4 m^2 \langle r^2 \rangle_{\rm nuc}
\label{06shabaevgl4}
\end{eqnarray}
for $ns$ states and
\begin{eqnarray}
  \Delta g = \frac{2(n^2-1)}{3n^5} (\alpha Z)^6 m^2 
\langle r^2 \rangle_{\rm nuc}
\end{eqnarray}
for $np_{\frac12}$ states. In the case of the $1s$ state,
the expression (\ref{06shabaevgl4})
coincides with the related formula in \cite{06karshenboim00a}.
A similar derivation for the nuclear-size correction to the
hyperfne spliting was performed in \cite{06volotka02}.

\begin{figure}    
\setlength{\unitlength}{0.5mm}\thinlines
  \begin{picture}(300,300)    
    
    \put(150,20){    
      \put(40,0){\line(0,1){100}}
      \put(40,30){     
  \multiput(0,0)(-10,10){3}{\oval(10,10)[bl]}    
  \multiput(-10,0)(-10,10){2}{\oval(10,10)[tr]}    
  \multiput(-10,40)(-10,-10){2}{\oval(10,10)[br]}    
  \multiput(0,40)(-10,-10){3}{\oval(10,10)[tl]}    }
     \put(40,50){    
\multiput(0,0)(2,0){10}{\line(1,0){1}}   
 \put(20,0){\circle*{3}} } 
}
      
    \put(70,20){    
      \put(40,0){\line(0,1){100}}
      \put(40,20){     
  \multiput(0,0)(-10,10){3}{\oval(10,10)[bl]}    
  \multiput(-10,0)(-10,10){2}{\oval(10,10)[tr]}    
  \multiput(-10,40)(-10,-10){2}{\oval(10,10)[br]}    
  \multiput(0,40)(-10,-10){3}{\oval(10,10)[tl]}    
}
     \put(40,85){    
\multiput(0,0)(2,0){10}{\line(1,0){1}}   
 \put(20,0){\circle*{3}} }
}    
    
    \put(-10,20){    
      \put(40,0){\line(0,1){100}}
     \put(40,40){     
  \multiput(0,0)(-10,10){3}{\oval(10,10)[bl]}    
  \multiput(-10,0)(-10,10){2}{\oval(10,10)[tr]}    
  \multiput(-10,40)(-10,-10){2}{\oval(10,10)[br]}    
  \multiput(0,40)(-10,-10){3}{\oval(10,10)[tl]}    
}
     \put(40,15){    
\multiput(0,0)(2,0){10}{\line(1,0){1}}   
 \put(20,0){\circle*{3}} }
}    

    \put(-10,160){    
      \put(40,0){\line(0,1){100}}
     \put(42,70){    
\multiput(0,0)(8,0){4}{\oval(4,4)[b]}    
\multiput(4,0)(8,0){3}{\oval(4,4)[t]}}    
     \put(78,70){\circle {19}}
     \put(40,30){    
\multiput(0,0)(2,0){10}{\line(1,0){1}}   
 \put(20,0){\circle*{3}} }
}    

    \put(70,160){    
      \put(40,0){\line(0,1){100}}
     \put(42,30){    
\multiput(0,0)(8,0){4}{\oval(4,4)[b]}    
\multiput(4,0)(8,0){3}{\oval(4,4)[t]}}    
     \put(78,30){\circle {19}}
     \put(40,70){    
\multiput(0,0)(2,0){10}{\line(1,0){1}}   
 \put(20,0){\circle*{3}} }
}    

    \put(150,160){    
      \put(40,0){\line(0,1){100}}
      \put(42,50){    
\multiput(0,0)(8,0){3}{\oval(4,4)[b]}    
\multiput(4,0)(8,0){2}{\oval(4,4)[t]}}    
     \put(70,50){\circle {19}}
     \put(80,50){    
\multiput(0,0)(2,0){6}{\line(1,0){1}}   
 \put(12,0){\circle*{3}} }
}    

\put(30,145) {a}
\put(110,145) {b}
\put(190,145) {c}

\put(30,5) {d}
\put(110,5) {e}
\put(190,5) {f}

\end{picture}    

\caption{The first-order QED corrections to the interaction
of the electron with a magnetic field.}    

\end{figure}
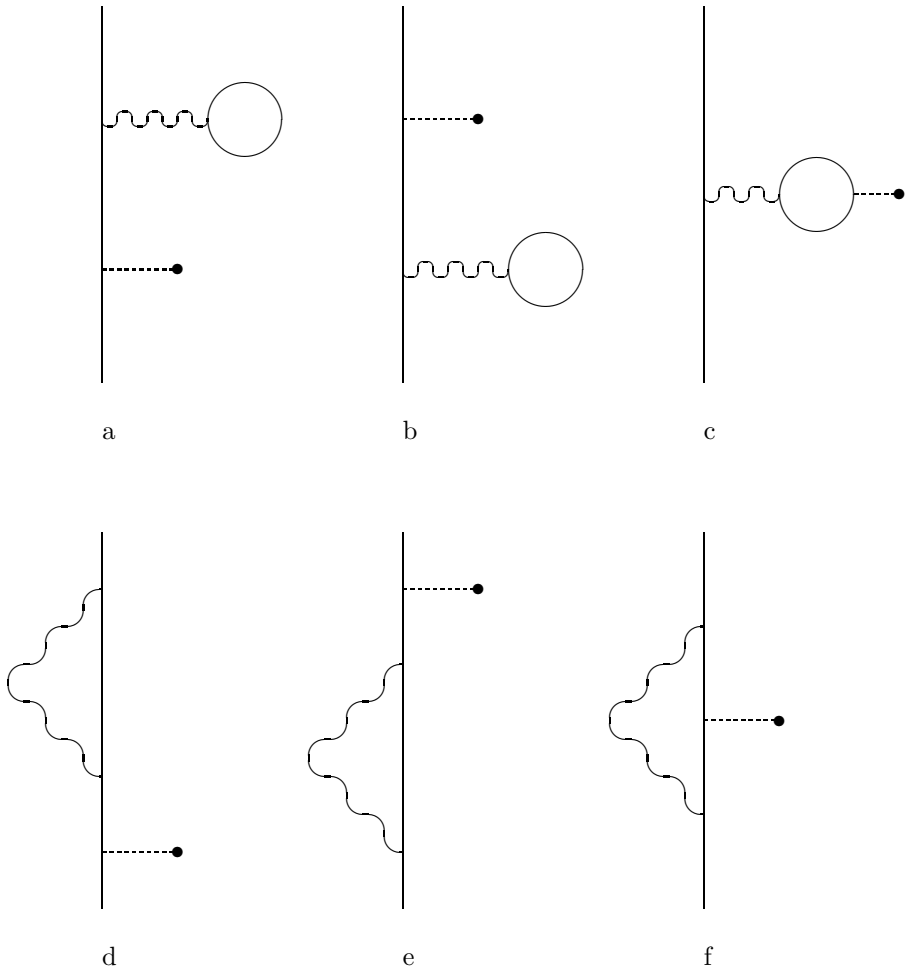

In \cite{06karshenboim01,06karshenboim00},
formulas (\ref{06gfac}) and (\ref{06hfs2})
were employed to evaluate analytically the one-loop 
vacuum-polarization (VP) corrections 
to the bound-electron $g$ factor and to the hyperfine splitting
in  low-$Z$ H-like atoms. These corrections are described
by diagrams presented in Fig. 1 (a-c), where the dashed line
ended by a circle indicates 
 the interaction with a magnetic field.
According to \cite{06karshenboim01}, for the $1s$ state,
 the VP correction to the $g$ factor
calculated in the Uehling approximation is given by
\begin{eqnarray}
\Delta g&=&\frac{\alpha}{\pi}\Bigl[-\frac{16}{15}(\alpha Z)^4
+\frac{5\pi}{9}(\alpha Z)^5 \nonumber\\
&&+\Bigl(
\frac{16}{15}\ln{(2\alpha Z)}-\frac{2012}{525}\Bigr)(\alpha Z)^6
\nonumber\\
&&+\Bigl(
-\frac{5\pi}{9}\ln{(\alpha Z/2)}-\frac{125\pi}{216}\Bigr)(\alpha Z)^7
 +\cdots\Bigr]\,.
\end{eqnarray}
The Uehling correction to the $1s$ hyperfine splitting is
\cite{06karshenboim00}
\begin{eqnarray}
\Delta E_{\rm hfs}&=&
\frac{\alpha}{\pi}E_{\rm F}\Bigl[\frac{3\pi}{4}(\alpha Z)
+\Bigl(\frac{34}{225}-\frac{8}{15}\ln{(2\alpha Z)}\Bigr)
(\alpha Z)^2
\nonumber\\
&&+
\Bigl(\frac{539\pi}{288}-\frac{13\pi}{24}\ln{(\alpha Z/2)}\Bigr)
(\alpha Z)^3 +\cdots\Bigr]\,,
\end{eqnarray}
where $E_{\rm F}$ is the Fermi energy.
In \cite{06yerokhin01,06yerokhin02},
 the one-loop self-energy corrections 
to the hyperfine splitting and to the bound-electron $g$ factor 
(Fig. 1 (d-f)) were calculated numerically to a high accuracy. 
Formulas (\ref{06gfac}) and (\ref{06hfs2}) were also employed in these
calculations.
In \cite{06nefiodov02}, formula (\ref{06gfac}) was used
to evaluate the nuclear-polarization effect on the bound-electron
$g$ factor.

\section{Other applications of the virial relations}

Applications of the virial relations are not restricted
only by H-like atoms and by the pure Coulomb field.
In \cite{06shabaeva95}, formulas (\ref{06hfs1}) and (\ref{06hfs2})
were employed to calculate the interelectronic-interaction
corrections to the hyperfine splitting in Li-like ions.
In \cite{06shabaev98}, virial relations 
(\ref{06bas1})-(\ref{06bas4}) with $V(r)\ne -\alpha Z/r$ 
were used to evaluate the recoil correction to 
the Lamb shift for an extended nucleus.
As an example, let us demonstrate how the virial relations
with $V(r)\ne -\alpha Z/r$  can be employed to evaluate
 the nuclear-size correction 
of the lowest order in $mR$ ($R$ is the nuclear charge radius)
to the integral
\begin{eqnarray}
C^{-2}=2\int_{0}^{\infty} dr\;r^{-2} G F\,
\end{eqnarray}
for an $ns$ state, which we denote by $|ns\rangle$.
This integral occurs in calculations of the hyperfine splitting.
We consider virial relations (\ref{06bas1})-(\ref{06bas4})
for the case $n=n'$, $\kappa=\kappa'$ and
for the potential
\begin{eqnarray} \label{06noncoul}
V(r)=-\frac{\alpha Z}{r} +\delta V(r)\,,
\end{eqnarray}
where $\delta V$ determines a deviation of the potential
from the pure Coulomb one due to the finite-nuclear-size effect.
Using notations (\ref{06defaa}), (\ref{06defbb}), and
(\ref{06defcc}), where $G$ and $F$ are calculated for
 potential   (\ref{06noncoul}), from equations
(\ref{06bas1})-(\ref{06bas4}) we derive
\begin{eqnarray}
&A^{-2}-2B^{-2}={\rm "regular\;\; terms"}\,, \\
&C^{-2}-2\langle ns|\sigma_z r^{-1} V|ns\rangle =
{\rm "regular\;\; terms"} \,,\\
&2A^{-2}-B^{-2}-2\langle ns|\sigma_x r^{-1} V|ns\rangle =
{\rm "regular\;\; terms"} \,,
\end{eqnarray}
where by "regular terms" we denote terms which have the 
nuclear-size corrections of order $(mR)^2$ and higher (it means that
their integrands  have more regular behaviour at
$r\rightarrow 0$ than the integrand of $C^{-2}$ has). 
From these equations, we easily obtain
\begin {eqnarray}
C^{-2}&=&\frac{2}{3-4(\alpha Z)^2}
\langle ns|
(3\sigma_z r^{-1}-2\alpha Z \sigma_xr^{-1})\delta V|ns\rangle\nonumber\\
&&+{\rm "regular\;\; terms"}\,.
\end{eqnarray}
To the lowest order in $mR$, it follows
\begin {eqnarray}
\delta C^{-2}=\frac{2}{3-4(\alpha Z)^2}
\langle ns|(3\sigma_z r^{-1}-2\alpha Z \sigma_xr^{-1})\delta V|ns\rangle\,.
\end{eqnarray}
An analytical evaluation of this expression 
to the two lowest orders in $\alpha Z$
yields \cite{06volotka}
\begin{eqnarray} \label{06del2}
\delta C^{-2}&=&
\frac{4}{n^3}(\alpha Z)^4
 m^3 \langle r\rangle_{\rm nuc}\Biggl\{1+(\alpha Z)^{2}
\Bigl[2\psi(3)-\psi(n+1)
\nonumber\\
&&-\frac{\langle r\ln{(2\alpha Zm r/n)}\rangle_{\rm nuc}}{\langle r
\rangle_{\rm nuc}}
+\frac{8n-9}{4n^2}+\frac{11}{4}\Bigr]\Biggr\}\,.
\end{eqnarray}
The non-relativistic limit is given by
\begin{eqnarray} \label{06del2nr}
\delta C^{-2}/C^{-2}=
-2\alpha Z m \langle r\rangle_{\rm nuc}\,.
\end{eqnarray}
Formula (\ref{06del2nr}) coincides with the related expression
derived in \cite{06shabaev94} for the sphere model for
the nuclear charge distribution, 
while
the relativistic $n$-independent
 term in formula (\ref{06del2}) differs
from the corresponding term that can be derived from
the formulas presented in \cite{06shabaev94}.
Since, for the sphere model,
 the approach developed in \cite{06shabaev94} provides a
more accurate evaluation of the nuclear size correction
 than the perturbation theory considered here,
formula (\ref{06del2}) can be improved
by replacing the relativistic $n$-independent term
with the corresponding term derived from \cite{06shabaev94}.
As a result, we obtain \cite{06volotka02}
\begin{eqnarray} \label{06del2m}
\delta C^{-2}&=&
\frac{4}{n^3}(\alpha Z)^4 m^3 \langle r\rangle_{\rm nuc}
\Biggl\{1+(\alpha Z)^{2}
\Bigl[2\psi(3)-\psi(n+1)
\nonumber\\
&&-\frac{\langle r\ln{(2\alpha Zm r/n)}
\rangle_{\rm nuc}}{\langle r\rangle_{\rm nuc}}
+\frac{8n-9}{4n^2}+\frac{839}{750}\Bigr]\Biggr\}\,.
\end{eqnarray}
Formulas (\ref{06del2}) and (\ref{06del2m}) differ
only by the last constant term.

The virial relations are also helpful for calculations
employing  finite basis set methods or analytical
expressions for the Coulomb-Green function. 
 In particular, they were
employed in \cite{06artemyev95,06shabaev98a}
 to calculate the nuclear recoil
corrections by using the B-spline method for the Dirac
equation \cite{06johnson88}. In that paper,
using the virial relations,
the original formulas for the recoil 
corrections, which contain some integrands with a singular behaviour
at $r\rightarrow 0$,  were expressed
in terms of less singular integrals. As a result,
the convergence of the numerical procedure for small $r$ 
was significantly improved. In \cite{06vrscay88},
the virial relations for  diagonal matrix elements
were used to construct Rayleigh-Schr\"odinger expansions
 for eigenvalues of perturbed radial Dirac equations to
arbitrary order.

\section{Conclusion}

In this paper we have considered the derivation of the virial
relations for the Dirac equation and their applications
for calculations of various physical quantities. 
It has been demonstrated that the virial relations are a 
very effective tool for analytical and high-precision numerical 
calculations of the hyperfine splitting and the bound-electron
$g$ factor in H-like ions. 
They are also useful for calculations
employing finite basis set methods and analytical expressions
for the Coulomb-Green function.

\section*{Acknowledgments}    
 
Valuable conversations with D.A. Glazov, V.G. Ivanov,
 U. Jentschura, S.G. Karshenboim,
 A.V. Nefiodov, A.V. Volotka,
and V.A. Yerokhin   
 are gratefully  acknowledged.    
This work was supported in part by RFBR (Grant No. 01-02-17248),
 by the program "Russian Universities" (Grant No. UR.01.01.072),
and by GSI.


    
\index{(Virial relations)}
\label{c_shab_}
\end{document}